\documentclass[prl,aps,twocolumn]{revtex4}

\usepackage{amsmath,graphicx,epsfig}
\usepackage{epstopdf}
\usepackage{euscript}
\usepackage{amsfonts}
\usepackage{amssymb}
\usepackage{float}

\def\prn#1{{\left(#1\right)}}

\def\bra#1{{\langle#1|}}

\def\pdbyd#1#2{{\frac{\partial#1}{\partial#2}}}

\def\cg(#1,#2)(#3,#4)(#5,#6){\bra{#1,#2,#3,#4}#5,#6\rangle}

\def\ts#1{{_{\mbox{\scriptsize #1}}}}

\def\threej(#1,#2)(#3,#4)(#5,#6){\begin{pmatrix}#1&#3&#5\\#2&#4&#6\end{pmatrix}}
\def\sixj(#1,#2,#3)(#4,#5,#6){\begin{Bmatrix}#1&#2&#3\\#4&#5&#6\end{Bmatrix}}
\def\ninej(#1,#2,#3)(#4,#5,#6)(#7,#8,#9){\begin{Bmatrix}#1&#2&#3\\#4&#5&#6\\#7&#8&#9\end{Bmatrix}}

\def\mb{\mathbf}

\newlength{\defbaselineskip}
\newcommand{\setlinespacing}[1]%
           {\setlength{\baselineskip}{#1 \defbaselineskip}}

\begin{document}

\title{In-situ measurement of light polarization with ellipticity-induced nonlinear magneto-optical rotation} 

\author{Derek F. Jackson Kimball}
\email{derek.jacksonkimball@csueastbay.edu}
\affiliation{Department of Physics, California State University --
East Bay, Hayward, California 94542-3084, USA}

\author{Jordan Dudley}
\affiliation{Department of Physics, California State University --
East Bay, Hayward, California 94542-3084, USA}

\author{Yan Li}
\affiliation{Department of Physics, California State University --
East Bay, Hayward, California 94542-3084, USA}

\author{Dilan Patel}
\affiliation{Department of Physics, California State University --
East Bay, Hayward, California 94542-3084, USA}

\date{\today}



\begin{abstract}
A precise, accurate, and relatively straightforward {\emph{in-situ}} method to measure and control the ellipticity of light resonantly interacting with an atomic vapor is described. The technique can be used to minimize vector light shifts. The method involves measurement of ellipticity-induced resonances in the magnetic-field dependence of nonlinear magneto-optical rotation of frequency-modulated light. The light propagation direction is orthogonal to the applied magnetic field $\mb{B}$ and the major axis of the light polarization ellipse is along $\mb{B}$. When the light modulation frequency matches the Larmor frequency, elliptically polarized light produces precessing atomic spin orientation transverse to $\mb{B}$ via synchronous optical pumping. The precessing spin orientation causes optical rotation oscillating at the Larmor frequency by modulating the atomic vapor's circular birefringence. Based on this technique's precision, {\emph{in-situ}} nature (which avoids systematic errors arising from optical interfaces), and independent control of the most important systematic errors, it is shown that the accuracy of light ellipticity measurements achievable with this technique can exceed that of existing methods by orders of magnitude.
\end{abstract}



\maketitle

Precise control of light polarization is crucial for many atomic physics experiments. Elliptically polarized light causes vector light shifts --- ac Stark shifts proportional to an atomic state's magnetic quantum number --- that affect atomic energy levels in a manner similar to that of a magnetic field \cite{Mat68,Bul71,Coh72,Hap72}. On the one hand, vector light shifts are useful, e.g., for manipulation of ultracold atomic gases \cite{Man03,Lin09,Sta13,Wan15} and for quantum information processing \cite{Bre99}, while on the other hand, vector light shifts cause troublesome systematic effects in atomic clocks \cite{Chi11,She12}, magnetometers \cite{Jen09,Bud13book,Ven07}, and precision tests of fundamental physics \cite{Rom99,Kim13,Swa13}. There have been a number of recent approaches developed to measure and control vector light shifts: for example, Zhu~{\it et al.}~\cite{Zhu13} used a variation of the Hanl\'e effect \cite{Kas73} with Cs atoms trapped in an optical lattice and Wood~{\it et al.}~\cite{Woo16} performed differential Ramsey interferometry on a pair of Bose-Einstein condensates. As noted in Refs.~\cite{Zhu13,Woo16}, although there are well-developed techniques for highly sensitive measurements of small polarization \emph{changes} \cite{Bir94,Dur10}, these techniques generally do not give absolute measurements of the polarization at similar levels --- they are typically limited to accuracies of $\sim 10^{-4}~{\rm rad}$ \cite{Bir94}. Here we present a sensitive, straightforward, {\emph{in-situ}} method to measure light ellipticity applicable to a diverse array of atomic physics experiments.

Our method employs nonlinear magneto-optical rotation (NMOR) resonances that arise when elliptically polarized light propagates through an atomic vapor. Experimental techniques involving NMOR \cite{Bud02review} are applied to diverse problems in magnetometry \cite{Bud07,Bud13book}, quantum and nonlinear optics \cite{Bud02review,Yas03,Mat03,Auz04,Pet04,Ott14}, and precision measurements \cite{Kim13,Vas09,Bro10,Gri09}. Our experiment takes advantage of narrow NMOR resonances related to long-lived, ground-state atomic spin polarization moments \cite{Kan95,Bud98,Bud00sensitivity,Bud02} which generically appear in atomic systems with slow relaxation of Zeeman coherences (long $T_2$) --- precisely those systems most significantly affected by vector light shifts. In our case, the long $T_2$ times are achieved by using an evacuated vapor cell whose inner surface is covered with an antirelaxation coating that enables atoms to bounce off the cell walls up to a million times while preserving their spin polarization \cite{Bal10}; similar effects should appear in any system with long $T_2$ times (for example, cold atoms trapped in far-detuned optical lattices \cite{Chi01,Blo08} and nitrogen-vacancy centers in diamonds \cite{Ken03,Bal09}). For the antirelaxation-coated rubidium (Rb) vapor cell used in these experiments, $T_2 \approx 320~{\rm ms}$ and is limited by spin-exchange collisions between Rb atoms \cite{Kim13}.

When the light-atom interaction is modulated at frequency $\Omega_m$, NMOR resonances related to different physical effects appear at different magnetic fields. In this work we employ NMOR of frequency-modulated light (FM NMOR) \cite{Bud02}, in which a single cw light beam is used for optical pumping and probing of the atomic spin polarization; analogous effects can be achieved in setups using amplitude-modulated light (AM NMOR) \cite{Gaw06,Pus08}. Note that superior sensitivity can often be achieved in two-beam arrangements where the pump and probe beam characteristics (power, detuning, etc.) can be separately optimized \cite{Pus06}. In experiments with modulated light, in addition to the zero-field NMOR resonance, resonances appear at magnetic fields where \cite{Yas03}:
\begin{align}
\Omega_m = \kappa \Omega_L~,
\label{Eq:FM-NMOR-resonance-condition}
\end{align}
where $\Omega_L$ is the Larmor frequency and $\kappa$ is an integer associated with the rank of the atomic spin polarization moment (PM) causing the optical rotation. The spatial distribution of angular momentum for the $q$ component of a PM has $|q|$-fold symmetry about the quantization axis, and the maximum possible $q$ is equal to the rank $\kappa$. For example, $\kappa = |q| = 1$ corresponds to orientation or a dipole moment (where the spin has a preferred direction or 1-fold symmetry) and $\kappa = |q| = 2$ corresponds to alignment or a quadrupole moment (where the spin has a preferred axis but no preferred direction, i.e., 2-fold symmetry). This explains the resonance condition: a PM returns to its initial configuration after a rotation of the spins by $2\pi/\kappa$, and so the condition for synchronous optical pumping of a PM of a particular $\kappa$ is given by Eq.~\eqref{Eq:FM-NMOR-resonance-condition}. In turn, the precessing PM modulates the optical properties of the medium at $\kappa \Omega_L$, leading to an observable FM NMOR signal (see detailed discussions in Refs.~\cite{Roc01,Bud13book}).

In the present work we investigate particular FM NMOR resonances generated by elliptically polarized light. The ellipticity-induced FM NMOR (EI FM NMOR) resonances provide an \emph{in-situ} method to directly measure the light ellipticity inside a vapor cell using the atoms themselves as the sensors. When light enters a vapor cell or passes through a window into a vacuum chamber, the polarization of the light is affected by the birefringence of the transparent material making up the cell wall or window. This causes initially linearly polarized light to become elliptically polarized, leading to uncontrolled vector light shifts that can adversely affect the accuracy of magnetometers \cite{Yab74,Nov01,Pat14} and other precision measurements based on spin precession \cite{Kim13,Swa13}. Note that elliptically polarized light has proven useful for eliminating dead zones in optical magnetometers \cite{Ben10} and for enhancing the amplitude of the zero-field NMOR resonance \cite{Nov01}.

\begin{figure}
\includegraphics[width=3.75 in]{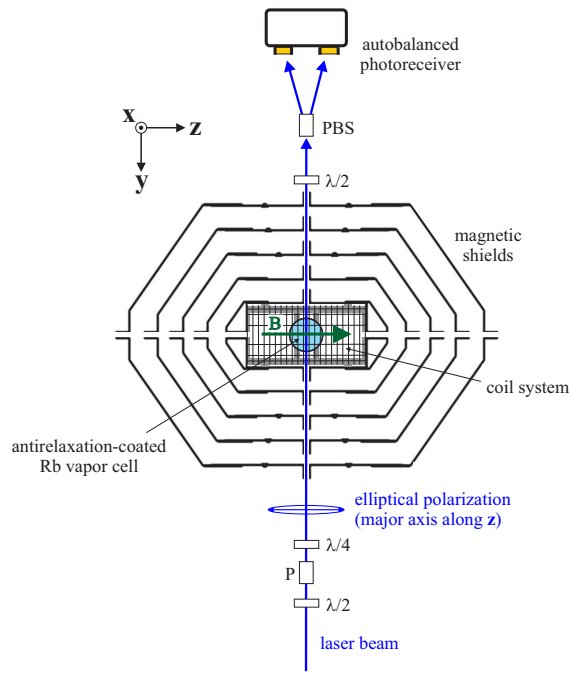}
\caption{Schematic of the experimental setup. P~=~linear polarizer, PBS~=~polarizing beamsplitter, $\lambda/4$~=~quarter-wave plate, $\lambda/2$~=~half-wave plate. Designation of $\mb{x}$, $\mb{y}$, and $\mb{z}$ directions is shown in the upper left corner. The green arrow at the center of the diagram represents the applied magnetic field $\mb{B}$. Assorted optics and electronics for laser control, data acquisition, and experiment control are not pictured.}
\label{Fig:setup}
\end{figure}

A simplified diagram depicting the experimental setup and geometry is shown in Fig.~\ref{Fig:setup} (further details concerning the experimental apparatus are given in Refs.~\cite{Kim09,Kim13}). An alkene-coated vapor cell (diameter =~5~cm), containing a natural isotopic mixture of Rb vapor (72\% $^{85}$Rb, 28\% $^{87}$Rb), is at the center of a system of coils nested within the innermost layer of a five-layer mu-metal shield \cite{Xu06,Kim16shielding}. The cell temperature was stabilized at $28.4^\circ{\rm C}$ corresponding to a Rb vapor density $\approx 2 \times 10^{10}~{\rm atoms/cm^3}$. The coils enable control of all three orthogonal components of the magnetic field and compensation of all first-order gradients. A magnetic field $\mb{B}$ is applied along the $z$~axis, perpendicular to the direction of propagation of the laser beam. Note that this geometry, the so-called Voigt geometry, is different from the Faraday geometry typically used in single-beam NMOR experiments. The major axis of the polarization ellipse of the light is aligned parallel to $\mb{B}$ (along $z$). The laser light, near-resonant with the Rb D2 transition ($^2S_{1/2}~\rightarrow~^2P_{3/2}$), is produced by a tunable external-cavity diode laser (Toptica DL100) and frequency modulated at $\Omega_m$ via sinusoidal laser diode current modulation. The laser beam diameter is $\approx 2~{\rm mm}$. Prior to entering the vapor cell, the probe beam passes through a $\lambda/2$ plate, a linear polarizer, and a $\lambda/4$ plate; these optical elements can be adjusted to control the input light's intensity and polarization, which are measured using a ThorLABs PAX720IR1-T polarimeter system. The polarimeter (not shown in Fig.~\ref{Fig:setup}) is inserted into and removed from the beam path as necessary. After exiting the antirelaxation-coated vapor cell, the beam is analyzed with a polarimeter consisting of a Wollaston prism polarizing beamsplitter whose output rays are detected with an autobalanced photoreceiver (New Focus Nirvana 2007).  The signal from the autobalanced photoreceiver is sent to the input of a digital lock-in amplifier (Signal Recovery model 7265) and demodulated at the first harmonic of $\Omega_m$.

The light ellipticity $\varepsilon$ is defined in terms of the normalized Stokes parameter $S_3$ (see, for example, Ref.~\cite{Auz10} for a detailed discussion):
\begin{align}
\varepsilon = \frac{1}{2} \arcsin S_3~.
\end{align}
The optical rotation angle $\varphi$ is defined in terms of the normalized Stokes parameters $S_1$ and $S_2$ \cite{Auz10}:
\begin{align}
\varphi = \frac{1}{2} \arctan \prn{\frac{S_2}{S_1}}~.
\end{align}
For our experimental geometry
\begin{align}
S_1 &= \prn{I_z - I_x}/I\ts{tot}~, \\
S_2 &= \prn{I_{+\pi/4} - I_{-\pi/4}}/I\ts{tot}~, \\
S_3 &= \prn{I_+ - I_-}/I\ts{tot}~,
\end{align}
where $I_z$ and $I_x$ are the light intensities along the $z$ and $x$ axes, $I_{+\pi/4}$ and $I_{-\pi/4}$ are the light intensities along axes tilted by $\pm \pi/4$ with respect to the $z$ axis, $I_+$ and $I_-$ are the left- and right-hand circular light intensities, respectively, and $I\ts{tot}$ is the total light intensity.

\begin{figure*}
\includegraphics[width=5 in]{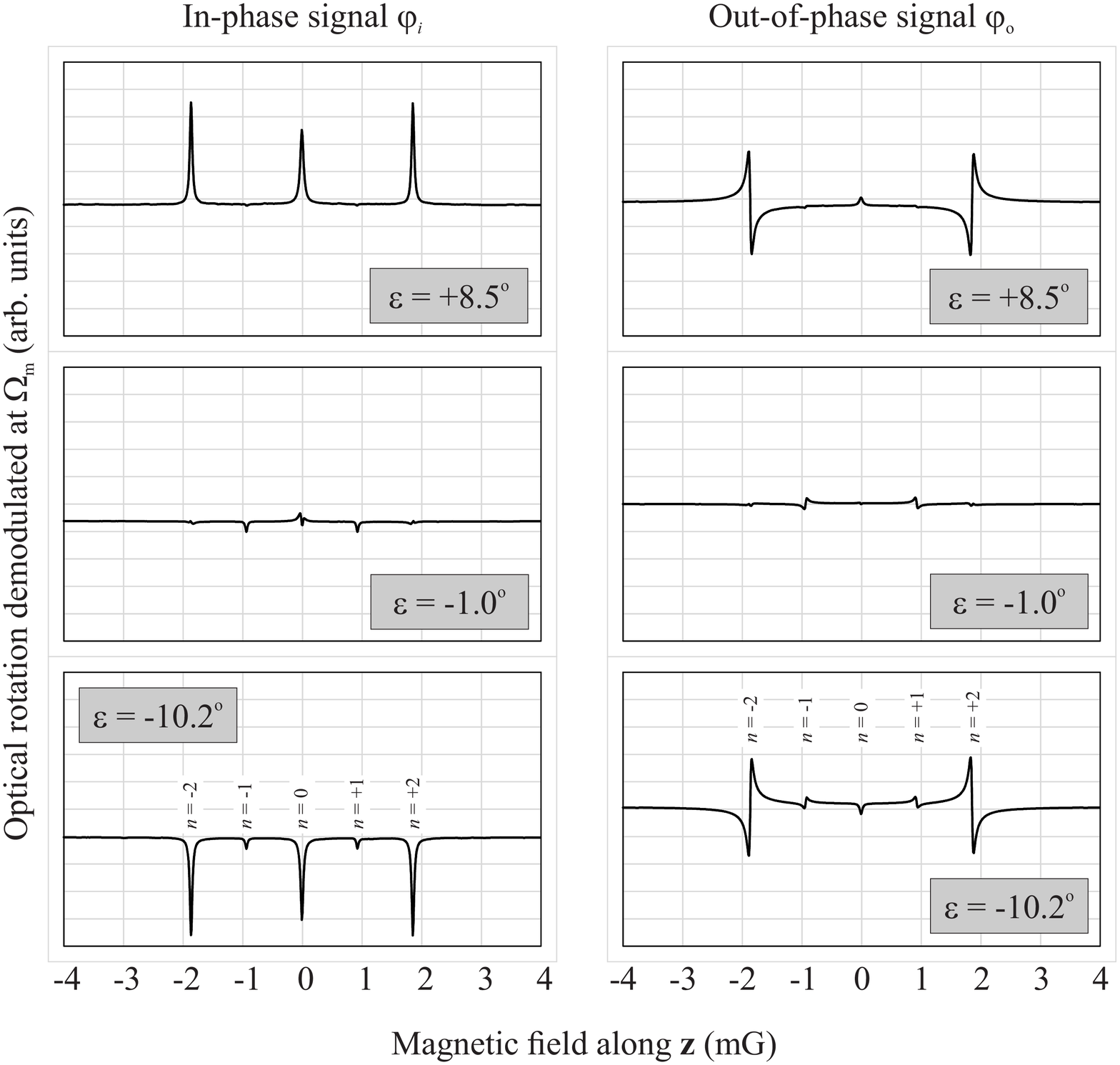}
\caption{Nonlinear magneto-optical rotation $\varphi$ as a function of applied magnetic field along the light polarization axis $z$ for different values of ellipticity $\varepsilon$, where $\varepsilon$ is measured in front of the vapor cell (prior to light entering the cell). The optical rotation signal is demodulated with a lock-in amplifier at the first harmonic of the laser modulation frequency $\Omega_m$. The plots on the left show the in-phase component $\varphi_i$ and the plots on the right show the out-of-phase component $\varphi_o$.  Light power $= 500~{\rm \mu W}$, laser tuned to the high-frequency wing of the $^{85}$Rb $F=3 \rightarrow F'$ transition, modulation frequency $\Omega_m = 2\pi \times 1~{\rm kHz}$, frequency-modulation amplitude $\Delta \omega = 2\pi \times 65~{\rm MHz}$. Magnetic resonances occur when $\Omega_L = \Omega_m/\kappa$, where the $n=\pm 2$ resonances correspond to the $\kappa=1$ condition and the $n=\pm 1$ resonances correspond to the $\kappa=2$ condition. The zero magnetic field resonance labeled $n=0$ generally has contributions from all $\kappa$. The $n=\pm 1$ resonances for the $\varepsilon = -1.0^\circ$ case, where the \emph{in-situ} ellipticity of the light is near zero, are due primarily to misalignment between the major axis of the polarization ellipse and the $z$-axis, leading to signals arising from precession of atomic alignment \cite{Pus06tilted} and alignment-to-orientation conversion \cite{Bud00aoc,Kim09}. Similar misalignment causes the appearance of $n=\pm 1$ resonances for the $\varepsilon = -10.2^\circ$ case as well. For this data, absolute calibration of the optical rotation angles was not carried out so the vertical axes are given in arbitrary units.}
\label{Fig:Bz-scans}
\end{figure*}

Light linearly polarized along $z$ optically pumps atomic spin alignment ($\kappa = 2$) along $\mb{B}$. Since the spin polarization is along $\mb{B}$, the magnetic torque on the atomic spins is zero; the spin polarization is static and does not precess. If the light is elliptically polarized, it optically pumps both alignment along $z$ and orientation ($\kappa = 1$) along $\pm\mb{k}$ (depending on its helicity), where $\mb{k}$ is the wave vector of the light. Spins oriented along $\pm\mb{k}$ experience a nonzero torque from $\mb{B}$ and precess. If the light-atom interaction is modulated synchronously with $\Omega_L$, macroscopic precessing orientation is generated. As the spin orientation precesses, the circular birefringence of the atomic medium is modulated, which in turn generates oscillating optical rotation. This results in EI FM NMOR resonances when $\Omega_m = \Omega_L$, the so-called $n = \pm 2$ resonances, as seen in Fig.~\ref{Fig:Bz-scans}. The plots on the left-hand side show the magnetic-field dependence of the amplitude of the oscillating optical rotation in-phase with the modulation of the light-atom interaction, labeled $\varphi_i$, and the plots on the right-hand side show the amplitude of the out-of-phase optical rotation, labeled $\varphi_o$ (see, for example, Refs.~\cite{Kim09,Mal04} for more details concerning the in-phase and out-of-phase optical rotation).

The choice of geometry is important for minimization of systematic effects. In the upper plots of Fig.~\ref{Fig:Bz-scans} that show the magnetic field dependence of optical rotation for ellipticity $\varepsilon = 8.5^\circ$, the major axis of the polarization ellipse is apparently well-aligned with $\mb{B}$. This is evident from the fact that the amplitudes of the $n=\pm 1$ FM NMOR resonances are relatively small. On the other hand, noticeable $n = \pm 1$ resonances appear in the lower plots of Fig.~\ref{Fig:Bz-scans} for $\varepsilon = -10.2^\circ$, indicating poor alignment of the major axis of the polarization ellipse with $\mb{B}$. The $n=\pm 1$ FM NMOR resonances correspond to the $\kappa=2$ resonance condition, $\Omega_L = \Omega_m/2$ which, as discussed above, is related to atomic spin alignment. Atomic spin alignment can precess if the major axis of the polarization ellipse is tilted with respect to $\mb{B}$, and in fact can generate both $n=\pm 1$ resonances and $n=\pm 2$ resonances depending on the tilt with respect to $\mb{k}$, as discussed in detail in Ref.~\cite{Pus06tilted}. Additionally, if optically pumped atomic alignment is tilted with respect to $\mb{B}$, the phenomenon of alignment-to-orientation conversion \cite{Bud00aoc,Kim09} can occur, where tensor light shifts coupled with Zeeman shifts induce more complicated evolution of the PMs \cite{Roc01}. Alignment-to-orientation can generate $n = \pm 2$ optical rotation signals that overlap with the EI FM NMOR resonances. Fortunately, the $n = \pm 1$ resonances themselves can be used to align the major axis of the polarization ellipse along $\mb{B}$ by making adjustments until their amplitude is minimized. To achieve pure linear polarization, this suggests an iterative process where the $n = \pm 1$ FM NMOR resonances are minimized to align $\mb{B}$ with the light polarization axis and the $n = \pm 2$ resonances are minimized to bring $\varepsilon$ to zero. The important role of light modulation should be emphasized: for unmodulated light (as used in Ref.~\cite{Mat03}), there is only a $n=0$ NMOR resonance to which PMs with different $\kappa$ all contribute, making their effects difficult to discriminate experimentally.

\begin{figure}
\includegraphics[width=3.5 in]{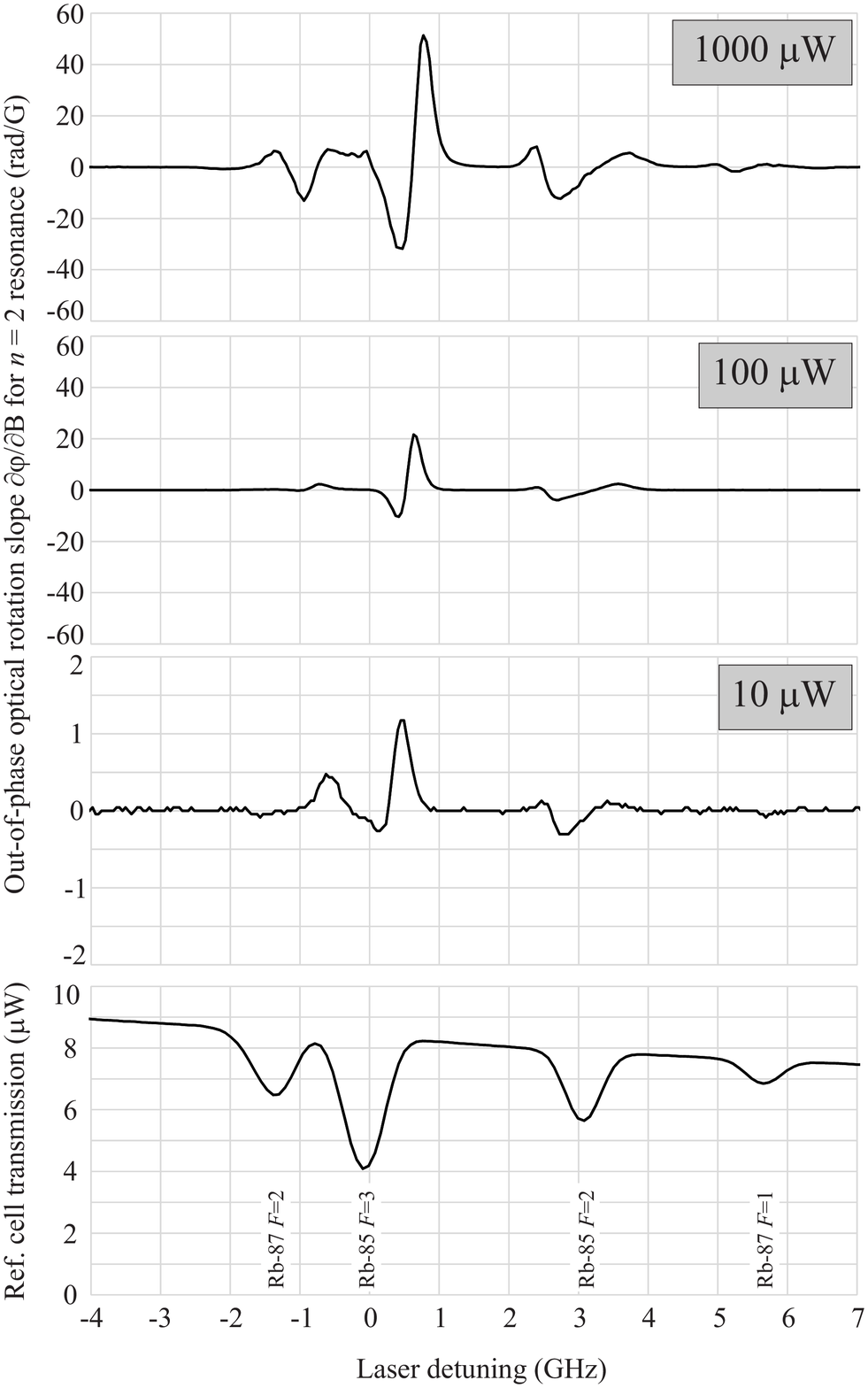}
\caption{Upper three plots show the slope ($\partial \varphi_o / \partial B$) of the out-of-phase component of the $n=2$ EI FM NMOR signal as a function of laser detuning for different incident light powers ($10~\mu W$, $100~\mu W$, and $1000~\mu W$, as indicated in the upper right corner of the plots). The lower plot shows the absorption spectrum of unmodulated light transmitted through a Rb reference cell. Light ellipticity as measured in front of the vapor cell is $\varepsilon = 8.3^\circ$, other conditions are the same as for the data shown in Fig.~\ref{Fig:Bz-scans}. Note the change in the scale of the vertical axis ($30\times$) between the plot for light power = $10~\mu W$ and the plots for $100~\mu W$ and $1000~\mu W$.}
\label{Fig:spectra}
\end{figure}


Figure~\ref{Fig:spectra} shows the spectral dependence of the $n=2$ EI FM NMOR signals for three representative light powers by plotting the slope of the out-of-phase component of the optical rotation with respect to magnetic field, $\partial \varphi_o / \partial B$, evaluated at $\Omega_L = \Omega_m$. Studying $\partial \varphi_o / \partial B$ helps eliminate the contribution of non-$B$-dependent background optical rotation to the spectra, for example that due to self-rotation of the elliptically polarized light \cite{Roc01self}. For reference, the transmission spectrum without laser frequency modulation is displayed at the bottom of Fig.~\ref{Fig:spectra} and the corresponding hyperfine components are labeled. At all light powers studied, the largest $\partial \varphi_o / \partial B$ is obtained when the laser light is detuned to the high frequency wing of the $^{85}$Rb $F=3 \rightarrow F'$ transition. Similar results have been found in past NMOR studies \cite{Bud98,Bud00sensitivity,Bud02,Kim09}, and are attributed to the role of the bright cycling transition, $F=3 \rightarrow F'=4$. When the light is resonant with the $F=3 \rightarrow F'=4$ transition, atoms are optically pumped into a bright state that interacts more strongly with the light. Another important factor causing the EI FM NMOR signal to be maximized on the slope of a Doppler-broadened optical resonance is that the $\Omega_m = \Omega_L$ resonance condition depends on modulation of the \emph{light-atom interaction}. If the laser is tuned to the center of a Doppler-broadened transition, the light-atom interaction is actually being modulated at $2\Omega_m$ which does not fulfill the resonance condition (see the detailed discussion of a similar situation in FM NMOR in Ref.~\cite{Kim09}).

\begin{figure}
\includegraphics[width=3.5 in]{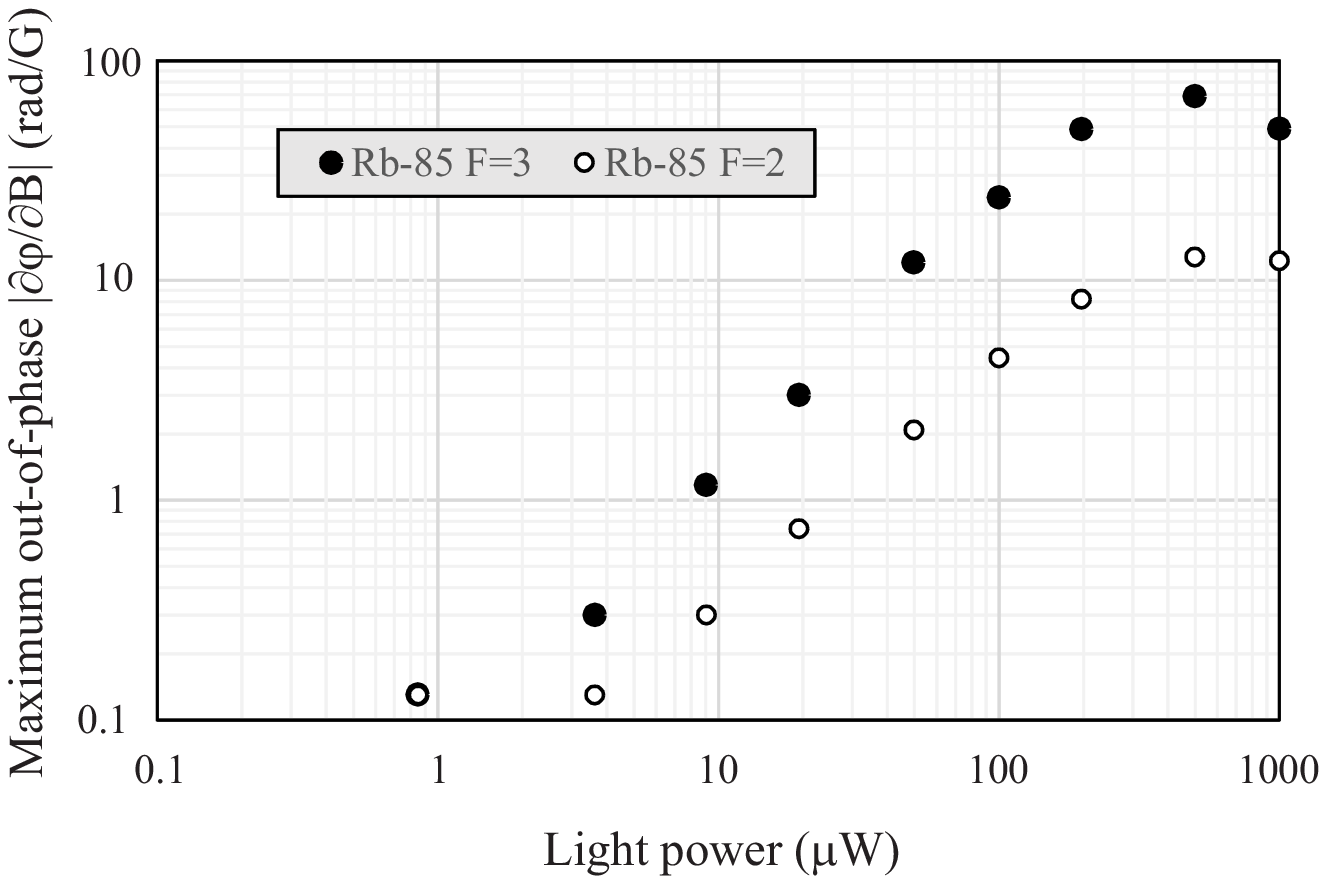}
\caption{The magnitude of the slope ($|\partial \varphi_o / \partial B|$) of the out-of-phase component of the $n=2$ optical rotation signal maximized with respect to laser detuning as a function of incident light power for the two $^{85}$Rb hyperfine components. Conditions are the same as for the data shown in Fig.~\ref{Fig:spectra}.}
\label{Fig:slope-vs-power}
\end{figure}

Figure~\ref{Fig:slope-vs-power} shows the detuning-maximized $\left| \partial \varphi_o / \partial B \right|$ as a function of incident light power for the $^{85}$Rb $F=3 \rightarrow F'$ (filled circles) and $^{85}$Rb $F=2 \rightarrow F'$ (unfilled circles) transitions. Note that the maximum $\left| \partial \varphi_o / \partial B \right|$ as a function of light power occurs near the same power for both transitions ($\approx 500~{\rm \mu W}$) and also that there is a monotonic increase up to the maximum power for both transitions. This is in contrast to the situation for linearly polarized light in the Faraday geometry \cite{Bud00aoc,Kim09}, where at sufficiently high light powers, alignment-to-orientation conversion becomes a dominant mechanism causing optical rotation, changing the sign of NMOR and FM NMOR effects and leading to non-monotonic behavior of the detuning-maximized rotation. This is in agreement with the expectation that due to the chosen geometry (major axis of the light polarization ellipse along $\mb{B}$), alignment-to-orientation conversion should not play any significant role in EI FM NMOR. The maximum of $\left| \partial \varphi_o / \partial B \right|$ as a function of light power can be explained by saturation effects related to optical pumping out of the probed ground-state hyperfine level into the unprobed ground-state hyperfine level (see discussions of related saturation phenomena in Refs.~\cite{Bud02review,Bud13book,Kim09,Bud00sensitivity}, where maxima occur at similar light powers).

\begin{figure}
\includegraphics[width=3.5 in]{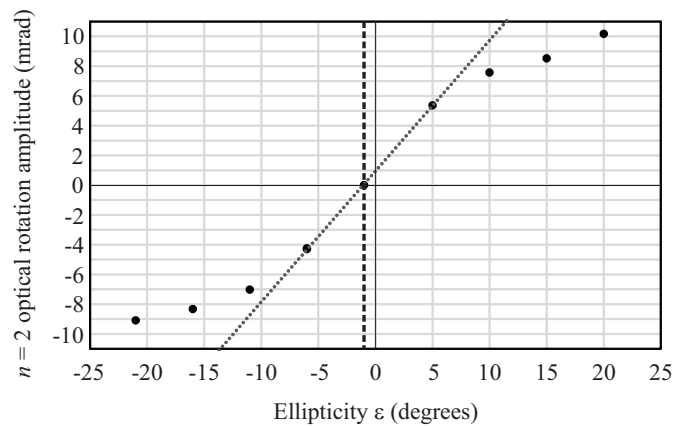}
\caption{The magnitude of the $n=2$ in-phase component of the magneto-optical rotation resonance as a function of ellipticity $\varepsilon$, where $\varepsilon$ is measured in front of the entrance to the vapor cell. Light power $= 500~{\rm \mu W}$, other conditions the same as for the data shown in Fig.~\ref{Fig:Bz-scans}. The gray dotted line shows the linear slope $\partial{\varphi_i}/\partial{\varepsilon}$ in the vicinity of the zero-crossing for $\varphi_i$. The thick black dashed line shows the offset of the $\varphi_i$ zero crossing of the optical rotation from $\varepsilon = 0$ is a result of the birefringence of the vapor cell walls, which induces ellipticity in nominally linearly polarized light. By zeroing the $n=2$ ellipticity-induced FM NMOR resonance, the value of $\varepsilon$ for light within the cell can be minimized.}
\label{Fig:rotation-vs-ellipticity}
\end{figure}

Figure~\ref{Fig:rotation-vs-ellipticity} shows the magnitude of the $n=2$ EI FM NMOR in-phase component as a function of $\varepsilon$ for an incident light power of $500~{\rm \mu W}$. The light is tuned to the high-frequency wing of the $^{85}$Rb $F=3 \rightarrow F'$ transition where EI FM NMOR is maximized. For sufficiently small $\varepsilon$, we find that
\begin{align}
\pdbyd{\varphi_i}{\varepsilon} = 8.7 \times 10^{-4}~{\rm rad/degree}~.
\end{align}
The standard quantum noise limit for polarimetry is
\begin{align}
\delta \varphi = \frac{1}{2\sqrt{N\ts{ph}}}~,
\label{Eq:photon-SNL}
\end{align}
where $N\ts{ph}$ is the number of photons per second detected by the polarimeter. For a light power of $500~{\rm{ \mu W}}$, Eq.~\eqref{Eq:photon-SNL} gives $\delta\varphi \approx 10^{-8}~{\rm rad}/\sqrt{\rm Hz}$. Thus we find the projected photon-shot-noise-limited sensitivity to ellipticity is
\begin{align}
\delta \varepsilon = \prn{ \pdbyd{\varphi}{\varepsilon} }^{-1} \delta \varphi \approx 2 \times 10^{-7}~{\rm \frac{rad}{\sqrt{Hz}}}~.
\end{align}
Shot-noise-limited polarimetry is routinely achieved in experiments using modulated light \cite{Bir94,Vas09,Bro10,Kim13}, and in our case our polarimeter is a factor of $\approx 3$ above the shot-noise limit \cite{Kim09,Kim13}.

This can be compared to the best existing measurement of ellipticity, that described in Ref.~\cite{Zhu13}, which is at the $10^{-5}~{\rm rad}$ level, but which involves the use ultracold atoms in an optical lattice. The EI FM NMOR technique described here is technically much simpler to implement and, if one averages for only $\approx 100~{\rm s}$, would represent an improvement in measurement sensitivity to $\varepsilon$ compared to Ref.~\cite{Zhu13} by roughly three orders of magnitude, illustrating the efficiency of EI FM NMOR techniques for ellipticity measurements.

One can also see the offset of the zero-crossing of the EI FM NMOR amplitude from $\varepsilon = 0$, where $\varepsilon$ was measured for the incident light prior to entering the vapor cell using the ThorLABs PAX720IR1-T polarimeter system. This indicates a systematic offset of $\varepsilon$ from nominal zero inside the cell, presumably caused by birefringence of the cell walls. This demonstrates the usefulness of EI FM NMOR as an \emph{in-situ} light polarization measurement technique. In fact, EI FM NMOR is used to zero the ellipticity of probe light in order to minimize vector light shifts in our ongoing experiment searching for a spin-gravity coupling \cite{Kim13}. We anticipate that this method will be similarly useful in other precision measurements of atomic spin precession.

\begin{figure}
\includegraphics[width=3.0 in]{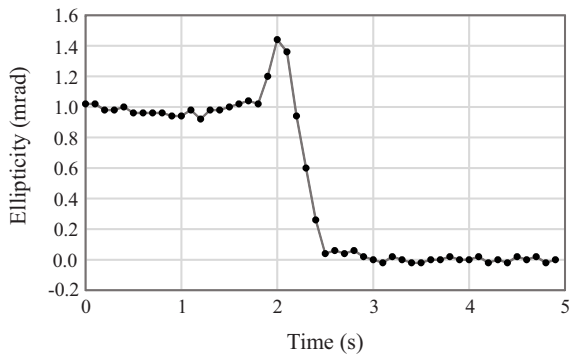}
\caption{Measurement of ellipticity $\varepsilon$ using EI FM NMOR. The ellipticity is changed by hand via rotation of a linear polarizer preceding a $\lambda/4$ plate at time $t \sim 2~{\rm s}$. Light power $= 100~{\rm \mu W}$, other conditions the same as in Fig.~\ref{Fig:Bz-scans}.}
\label{Fig:ellipticity-step}
\end{figure}

Figure~\ref{Fig:ellipticity-step} shows a measurement of ellipticity $\varepsilon$ as determined by the calibration shown in Fig.~\ref{Fig:rotation-vs-ellipticity}, where now $\varepsilon = 0$ is taken to be the zero crossing of $\varphi_i$. For the data shown in Fig.~\ref{Fig:ellipticity-step}, the light polarization is adjusted by hand to zero the ellipticity by rotating the linear polarizer preceding the $\lambda/4$ plate before the light enters the vapor cell (see Fig.~\ref{Fig:setup}). This is the final stage of an iterative process using both the $n=1$ and $n=2$ EI FM NMOR resonances to align the major axis of the polarization ellipse along $\mb{B}$ by minimizing the $n=1$ resonance amplitudes and zeroing the ellipticity by minimizing the $n=2$ resonances. The measured ellipticity, averaged over the last two seconds of data shown in Fig.~\ref{Fig:ellipticity-step}, has an uncertainty of $3.4 \times 10^{-6}~{\rm rad}$, demonstrating the precision of the method. The precision of the measurement is lower than the shot-noise limit for two reasons: (1) as mentioned above, our polarimeter operates a factor of $\approx 3$ away from the shot-noise limit due to technical noise, and (2) the light power used in this measurement is $100~{\rm \mu W}$ rather than $500~{\rm \mu W}$, so $\varphi_i$ is a factor of $\approx 3.5$ away from its maximum value.

The accuracy with which $\varepsilon$ can be measured is evaluated by consideration of possible systematic errors. Systematic errors due to birefringence of optical elements (such as the vapor cell walls) before and after the atomic sample are negligible since the EI FM NMOR signal is produced by the atoms themselves and has a sharp resonant character in optical frequency, modulation frequency, and magnetic field. As described above, for our geometry where the major axis of the polarization ellipse is along $\mb{B}$, the symmetry of the system prevents any appearance of $n = \pm 2$ resonances occurring at $\Omega_m = \Omega_L$ unless $\varepsilon \neq 0$. Thus we conclude in principle and observe in practice (see Fig.~\ref{Fig:Bz-scans}) that the dominant systematic error comes from deviation from this ideal geometry: if the major axis of the polarization ellipse is rotated by an angle $\theta$ with respect to $\mb{B}$ then we observe $n = \pm 2$ FM NMOR resonances even when $\varepsilon=0$, as discussed in detail in Ref.~\cite{Pus06tilted}. However, as noted above, FM NMOR offers a useful tool with which to independently measure and minimize $\theta$, namely the $n = \pm 1$ FM NMOR resonances. Again from symmetry considerations, to first order elliptically polarized light does not affect the amplitude of the $n = \pm 1$ FM NMOR resonances since they occur at $\Omega_m = 2\Omega_L$ and are related to $\kappa=2$ atomic spin alignment. If the major axis of the polarization ellipse is aligned along $\mb{B}$, atomic spin alignment does not precess and thus cannot generate the $n = \pm 1$ FM NMOR resonances. Note that in this geometry, alignment-to-orientation effects are also zeroed \cite{Bud00aoc,Kim09}. By zeroing the amplitude of the $n = \pm 1$ FM NMOR resonances, the misalignment angle $\theta$ can also be zeroed and the primary systematic error can be controlled. The amplitudes of both the non-ellipticity-induced $n = \pm 2$ resonances and the $n = \pm 1$ resonances scale as $\theta^2$ for $\theta \ll 1$. Using the techniques described in Ref.~\cite{Pus06tilted} we can tune $\theta \lesssim 10^{-4}~{\rm rad}$, which generates a $n = \pm 2$ FM NMOR resonance amplitude of $\approx \pm 4 \times 10^{-11}~{\rm rad}$ under our experimental conditions. This translates to a systematic uncertainty in $\varepsilon$ of $\approx \pm 10^{-9}~{\rm rad}$. This is far below the statistical uncertainty of the measurement shown in Fig.~\ref{Fig:ellipticity-step} and supports the conclusion that the measurement is statistically limited. After minimization of $\varepsilon$ using this procedure, measurements of vector light shifts in this system are consistent with $\varepsilon \lesssim 2 \times 10^{-4}$ as discussed in Ref.~\cite{Kim13}.

In considering the application of EI FM NMOR to other systems of interest (such as cold atoms and nitrogen-vacancy centers in diamonds), we note that since our measurement is statistics-limited, the sensitivity to $\varepsilon$ scales as $1/\sqrt{NT_2}$, where $N$ is the number of atoms (in our case $N \approx 10^{12}~{\rm atoms}$). Another point to be noted is that in our experiment, because the Rb atoms are contained in an evacuated, antirelaxation-coated cell, the quantity measured is in fact the average ellipticity across the light beam, since the atoms are not spatially localized \cite{Zhi16}. Such motional averaging is reduced in the case of cold atoms, nitrogen-vacancy centers in diamonds, and vapor cells filled with buffer gas. In these cases it should be possible to use EI FM NMOR to study the spatial profile of $\varepsilon$.

In conclusion, we have introduced and characterized a new \emph{in-situ} method to measure and control the ellipticity of light interacting with atomic vapors: ellipticity-induced nonlinear magneto-optical rotation of frequency modulated light (EI FM NMOR). This relatively straightforward method can be applied in experiments, for example, where precise and accurate control of vector light shifts is important.

\acknowledgments

The authors are sincerely grateful to Dmitry Budker, Szymon Pustelny, and Wenhao Li for invaluable discussions, to Mohammad Ali for technical work on parts of the apparatus, and to generations of students who worked on earlier iterations of the experimental apparatus, in particular Rene Jacome, Ian Lacey, Jerlyn Swiatlowski, and Julian Valdez.  This work was supported by the National Science Foundation under grants PHY-0652824, PHY-0969666, and PHY-1307507.  The findings expressed in this material are those of the authors and do not necessarily reflect those of the NSF.

\bibliography{EI-FM-NMOR-bib}

\end{document}